\documentclass[12pt,aps,pra,superscriptaddress]{revtex4}

\usepackage{amsmath}
\usepackage{amsfonts}
\usepackage{amssymb}
\usepackage{multirow}
\usepackage{verbatim}
\usepackage{alltt}
\usepackage{moreverb}
\usepackage{graphicx,color,graphics}
\usepackage{hyperref}
\usepackage{setspace}
\usepackage{url}

\usepackage{dcolumn}
\usepackage{bm}
\usepackage{amsthm}
\usepackage{textcomp}
\usepackage{amssymb}
\usepackage{xspace}
\usepackage{subfigure}

\newtheorem{theorem}{Theorem}

\newcommand{\qvbar}{{|}}
\newcommand{\qrangle}{{\rangle}}

\newcommand{\ket}[1]{\qvbar{#1}\qrangle}

\providecommand{\ignore}[1]{}
\providecommand{\aucmnt}[1]{#1}
\renewcommand{\aucmnt}[1]{}

\newcommand{\argmin}{\operatornamewithlimits{argmin}}

\begin{document}

\title{Optimizing Passive Quantum Clocks}
\author{Michael Mullan}
\affiliation{Department of Physics, University of Colorado at Boulder, Boulder, Colorado, 80309, USA}
\affiliation{National Institute of Standards and Technology, Boulder, Colorado, 80305, USA}
\author{Emanuel Knill}
\affiliation{National Institute of Standards and Technology, Boulder, Colorado, 80305, USA}

\begin{abstract}
  We describe protocols for passive atomic clocks based on quantum
  interrogation of the atoms.  Unlike previous techniques, our
  protocols are adaptive and take advantage of prior information
  about the clock's state.  To reduce deviations from an ideal clock,
  each interrogation is optimized by means of a semidefinite program
  for atomic state preparation and measurement whose objective
  function depends on the prior information.  Our knowledge of the
  clock's state is maintained according to a Bayesian model that accounts for
  noise and measurement results.  We implement a full simulation of a
  running clock with power-law noise models and find significant
  improvements by applying our techniques.
\end{abstract}

\maketitle

\section{Introduction}

Atomic clocks continue to make great strides in accuracy and
stability.  Passive atomic clocks compare the frequency of an
external, ``flywheel'' oscillator to that of a reference transition in
an atom, the atomic standard.  In view of the increasing impact of
quantum information science and the associated rapid growth of quantum
control capabilities, there has been substantial interest in the
possibility of exploiting quantum effects and quantum algorithms for
further improvements in clock precision.  Whereas measurements with
$N$ atoms in independent states yield errors that scale as
$1/\sqrt{N}$ (the so called ``standard quantum limit'', SQL),
protocols using entangled quantum states can yield errors that scale
as $1/N$, the fundamental ``Heisenberg limit''.  Here, we are
interested in optimizing such quantum protocols and evaluating their
performance in clocks subject to realistic noise.

The first proposals to beat the SQL used spin squeezed
states~\cite{squeeze3, squeeze, squeeze2}.  Not long afterward,
Bollinger et. al.~\cite{bollinger:qc1996a} demonstrated that the
Heisenberg bound could be achieved by using maximally entangled
states.  Even though such states achieve optimal scaling in the
noiseless regime, it was shown~\cite{huelga} that such states cannot
beat the SQL in the presence of atomic decoherence.  This spurred
research into protocols that perform well even in noisy systems; for
example, see Refs.~\cite{noisyProtocol,noisyProtocol2,
  noisyProtocol3,noisyProtocol4,noisyProtocol5,noisyProtocol6}.
However, in most modern clocks, the dominant sources of noise are
random fluctuations of the external oscillator (see below) and not
atomic decoherence.  Refs.~\cite{winelandBible,squeezedClocks} were
among the first to study clock optimization in the presence of this
type of noise. In Ref.~\cite{squeezedClocks}, an error scaling of
$1/N^{2/3}$ is obtained by optimizing over a family of spin squeezed
states.

The standard approach to the clock optimization problem involves
optimizing individual measurements of the atomic standard with respect
to fixed objective functions.  In this spirit, Refs.~\cite{buzek,
  dobrzanski, van2007optimal, macieszczak2013optimal} derive states
and measurements that are optimal under certain sets of assumptions by
optimizing a cost function that approximates a clock's performance.
In Ref.~\cite{numericalTest}, various such measurement protocols are
compared in a Monte Carlo simulation of a clock subject to $1/\omega$
noise.  For a two qubit clock, it was estimated that the best protocol
would result in a $15$-$20\,\%$ improvement in Allan variance, a standard
measure of long-term clock performance. However, none of the simulated
protocols were optimal, so despite this work and other work in quantum
metrology (for a review, see~\cite{giovannetti:qc2011a}), it is yet to
be seen how much can be gained by fully utilizing quantum resources.

Clocks are used to construct a time scale by marking, or timestamping,
a set of events labeled $\{1,2,\hdots n\}$ with time values $\{t_1,
t_2 \hdots t_n\}$.  Time is defined in terms of a transition frequency
$\Omega$ of a chosen atomic frequency standard; if the standard could
be measured directly, these assignments could be made trivially by
counting cycles of the selected transition.  However, in practice,
such a measurement is often difficult, and therefore, an external
oscillator or ``flywheel'' at frequency $\omega$ near $\Omega$ is
measured rather than the standard. Clocks that use such an oscillator
are referred to as ``passive'' atomic clocks.  A measurement of the
phase deviation of the external oscillator is then required to assign
timestamps accurately.  Such a measurement involves interactions
between the external oscillator and the standard.  The part of the
protocol involving one state preparation followed by evolution and a
measurement is called an interrogation.  Our goal here, and the goal
of the work discussed above, is to optimize passive clocks by deriving
protocols that maximize the information gained during each
interrogation.

Here we consider interrogations involving a state preparation, a free
evolution and a measurement, where the state preparation and
measurement take negligible time.  We gain information about the
time-averaged frequencies during interrogations.  Because the
interrogations necessarily yield incomplete information and the
external oscillator is subject to noise, the frequencies are described
by probability distributions.  Given a model of the noise affecting
our systems, our knowledge of the state of the system after $n$
interrogations is described by the probability distribution $p(
\langle \omega \rangle_1, \langle \omega \rangle_2, \hdots \langle
\omega \rangle_n | a_1, a_2, \hdots a_n )$, where $\langle \omega
\rangle_k$ is the time-averaged frequency of the external oscillator
during interrogation $k$ and $a_k$ is the measurement outcome obtained
at the end of the $k$'th interrogation. To avoid confusion between
time-averaging for a particular instance of the noise model and
computing expected values based on the noise model's probability
distribution, we drop the averaging brackets and identify $\omega_k$
with the time-averaged frequency during the $k$'th interrogation.
Note that the $\omega_k$ are random variables that can be expressed as
integrals over the instantaneous frequencies.  We abbreviate their
distributions as $p( \omega_1, \omega_2, \hdots \omega_n | a_1, a_2,
\hdots a_n ) = p( \boldsymbol{ \omega_n} | \boldsymbol{a_n})$, where
$\boldsymbol{\omega_n}$ and $\boldsymbol{a_{n}}$ refer to the sequence
of time-averaged frequencies and the sequence of measurement outcomes
obtained during interrogations $1$ through $n$.  Unless required for
clarity, we drop the adjective ``time-averaged'' when referring to the
$\omega_k$.

Our approach improves on prior work in several ways.  First, observe
that the optimal choice for the $n+1$'th interrogation depends on $p(
\boldsymbol{ \omega_n} | \boldsymbol{a_n})$.  Given a good model of
the noise, it is possible to keep track of these conditional
probability distributions.  Traditional interrogations do not take
advantage of this information; many use the same, fixed strategy for
each interrogation.  Furthermore, most analyses related to the
Heisenberg limit apply only in the absence of preexisting information.
In contrast, our interrogations are dynamic. They are tailored to our
knowledge of the clock's state by making use of available prior
information.  Second, we jointly derive quantum state preparations and
measurements with a semidefinite program. We refer to the quantum
state preparation and measurement as the quantum algorithm used by the
interrogation. The semidefinite program gives us freedom in choosing
our optimization criteria, which are expressed in the form of
state-dependent cost functions.  Given such a function, the
semidefinite program determines the optimal quantum algorithm without
being limited to specific cost functions or a restrictive class of
states and measurements. Third, we prove that this flexibility in
choosing cost functions is required in order to minimize the error in
the total elapsed-time estimates. For most noise models, memory
effects imply that a simple criterion based on the difference between
the current frequency of the oscillator and the estimated one does not
suffice.

The remainder of this paper is structured as follows: We discuss
atomic interrogation and detail the properties of the types of noise
assumed to affect the external oscillator.  We then describe our
optimization criteria and explain how we dynamically derive the
quantum algorithm for the next interrogation.  Finally, we implement
our protocol on full simulations of clocks subject to various
power-law noise models and demonstrate improvements over prior, fixed
protocols.  Throughout, we assume full quantum control over the atomic
system and that the only source of noise is statistical fluctuations
of the external oscillator.  While this latter assumption is sensible
for many modern clocks, if necessary, our scheme can be adapted to
account for decoherence~\cite{firstClock}. We conclude with a
discussion of further work needed to apply our theoretical methods to
experimental passive clocks.

\section{Interrogations}

Here, we consider the atoms as idealized two-level systems with
standard basis $\ket{0}$ and $\ket{1}$ and use the usual conventions
for operators acting on these systems.  Once a quantum algorithm has
been decided on, an interrogation prepares the atoms in an initial
state $\ket{\psi}$ via the application of a chosen unitary,
$\ket{\psi} = U\ket{\mathbf{0}}$ to the standard starting state
$\ket{\mathbf{0}}$.  The inital state preparation is followed by a
period of free evolution of duration $T$, which effects a $z$ rotation
$R_z((\omega -\Omega)T)$ by an angle of $(\omega - \Omega)T$. Thus we
transform $\ket{\psi}$ to $\ket{\psi '} = R_z(( \omega -
\Omega)T)U\ket{\textbf{0}}$.  The angle of the rotation relates the
external oscillator's frequency to that of the atomic standard.  We
assume that the time needed to apply unitaries is negligible compared to
the period of free evolution.  Afterward, the atomic state $\ket{\psi
  '}$ is measured with a complete positive operator-valued measure
(POVM) $\{P_a\}_a$. Traditional interrogations choose the same initial
state and, except for a phase, the same POVM every time.  For example,
the widely used Ramsey method prepares $N$ atoms in the state
$\left( \begin{smallmatrix} 1/ \sqrt{2} \\
    -i/\sqrt{2} \end{smallmatrix} \right)^{\otimes N} $ and, after a
period of free evolution, measures each atom independently in the
$\ket{+}$, $\ket{-}$ basis.  Here we allow $U$ and the POVM to be
chosen differently in each interrogation.  Note that the quantum
algorithm of an interrogation can be generalized to $\ket{\psi '} =
R_z(( \omega - \Omega)T) U(t_f) \hdots U(1) R_z(( \omega - \Omega)T)
U(0) \ket{\mathbf{0}}$, followed by a measurement of $\ket{\psi '}$
with a POVM.  These multi-round algorithms can outperform single-round
ones.  However, they are more difficult to implement in
practice. Although our optimization procedures can derive such
algorithms, we do not consider them here.

Since all measurements will be referenced to $\Omega$, from now on we
take $\omega$ to be the frequency deviation from $\Omega$ rather than
the absolute frequency. We normally omit the modifier ``deviation''.

\section{Noise Models}

Noise affects the external oscillator at all times, competing with the
knowledge gained from measurements.  The noise model determines the
prior distributions to be used for the frequencies. Here, we assume
that it can be approximated by a continuous, multivariate Gaussian
random process characterized by a spectral density, $S(\omega)$.  It
has been determined that power law noise is a good approximation on
relevant frequency ranges, in which case $S(\omega) \propto
\omega^{\alpha}$. Relevant exponents are $\alpha \in
[-4,2]$~\cite{allan,riley:qc2008a}.  For example, $1/\omega$ noise is
common for cavity-locked optical
oscillators~\cite{optical1f}. Gaussian noise processes are
characterized by their means and covariances.  For our applications,
the unconditional means are assumed to be zero.  The $\omega_k$ can
then be characterized as joint Gaussian random variables characterized
by their covariances.  We denote the covariance between the external
oscillator's frequencies at times $s$ and $s'$ as
$\mathbf{Cov}(\omega(s), \omega(s'))$. Strictly speaking, $\omega$ and
the covariances need to be interpreted as generalized functions of
time. As we define them below, their domain is restricted to test
functions with zero mean. In particular, we compute covariances only
for differences between interval averages. For point values or for
other averages, the expressions given may be undefined or fail to give
non-negative variances.  We focus on frequency changes relative to an
initial frequency, where for the purpose of defining our priors, the
initial frequency is taken to be zero. With respect to the
experimentally relevant frequencies, we define the covariance matrix
$C$ according to
\begin{equation}
\begin{array}[b]{rl@{}l}\displaystyle
C_{i,j} &= \mathbf{Cov}(&\omega_i-\omega_0,\omega_j-\omega_0) \\
 &= \mathbf{Cov}(& \langle \omega(s) \rangle_{s\in I_{i}} -\langle \omega(s) \rangle_{s\in I_0},  \\
 &&\langle \omega(s) \rangle_{s\in I_{j}} - \langle \omega(s) \rangle_{s\in I_0} ),
\end{array}
\label{eq:aveCovs}
\end{equation}
where $I_i$ is the $i$'th interrogation interval
$I_i = [t_i,t_{i+1}]$, and $I_0=[t_0,t_1]$ is an interval before the
first interrogation. The length of the $i$'th interval is defined
as $T_i$. 

To compute $C_{i,j}$ for $-3<\alpha<-1$ we can formally express
$\mathbf{Cov}(\omega(s),\omega(s')) =
-\frac{h_\alpha}{2}|s-s'|^{-\alpha-1}$ where $h_{\alpha}$ is an
$\alpha$ dependent scale factor.  For $\alpha=-1$,
$\mathbf{Cov}(\omega(s),\omega(s')) = -2h_{-1}\ln|s-s'|$~\cite{1f}.
The $C_{i,j}$ can then be computed by expanding to a sum of terms of
the form
\begin{eqnarray}
\mathbf{Cov}(\langle \omega(s)\rangle_{s\in I_k},\langle \omega(s)\rangle_{s\in I_l}) \hspace*{-1.5in}&&\nonumber\\
&=& 
 \frac{1}{T_k}\frac{1}{T_l}
 \int_{t_{k}}^{t_{k+1}}ds\int_{t_l}^{t_{l+1}}ds' \mathbf{Cov}(\omega(s),\omega(s')).
\end{eqnarray}
For $\alpha\leq -3$, one can view the power spectrum as a $2r$'th
distributional derivative of a power spectrum with $\alpha\in(-3,-1]$
and multiply the formal expression for
$\mathbf{Cov}(\omega(s),\omega(s'))$ by $(-1)^r(s-s')^{2r}$ to apply
similar techniques.  For $\alpha > -1$, the distributional derivatives
are applied to $\mathbf{Cov}(\omega(s),\omega(s'))$.  In both cases,
care must be taken to ensure that covariances are computed only for
quantities in the appropriate domain where they are well-defined and
positive-definite.  

\section{Optimizing Interrogations}

The quality of our true-time estimates depends on how well we estimate
the phase difference between the external oscillator and an ideal
oscillator with frequency that of the atomic standard.  We therefore
wish to choose phase estimates, $\theta^{*}_n$, that minimize the
expectation 
\begin{equation}
\label{eq:error}
\mathbf{E}\left( \left( \theta_n - \theta^{*}_n \right)^2 \right),
\end{equation}
where $\theta_n$ is the cumulative phase difference of the external
oscillator after interrogation $n$, $\theta_n = \omega_1 T_1 +\omega_2
T_2 \hdots + \omega_n T_n$, and $\theta^{*}_n$ is our estimate of this
phase.  The expectation is taken over the noise model and we use the
symbol $\mathbf{E}$ to denote the expectation.  The expression
in~\eqref{eq:error} is evaluated according to
\begin{equation}
\label{eq:error2}
\mathbf{E}\left( \left( \theta_n - \theta^{*}_n \right)^2 \right) = 
\int (\theta_n - \theta_n^*)^2 p(\theta_n | \boldsymbol{a_n} )d\theta_n,
\end{equation}
where $p(\theta_n | \boldsymbol{a_n} )$ can be obtained from $p(
\boldsymbol{\omega_n} | \boldsymbol{a_n} )$.  The choice $\theta^{*}_n
= \mathbf{E}( \theta_n | \boldsymbol{a_n} )$ minimizes
Eq.~\eqref{eq:error2}, giving a value equal to the posterior variance,
$V_n=\mathbf{V}( \theta_n |\boldsymbol{a_n} )$.  Here, we use the
symbol $\mathbf{V}$ to denote the variance.  Our goal is therefore to
construct quantum algorithms that minimize the expected posterior
variance increase $\Delta V_n$ given by
\begin{align}
\label{eq:criteria}
\Delta V_{n} &= \sum_{a_{n}} (V_{n}-V_{n-1}) p(a_n | a_1 \hdots a_{n-1} )\notag\\
&=\sum_{a_n} \mathbf{V}( \theta_{n-1} + \omega_{n}T_{n} | a_1 \hdots a_n ) p(a_n | a_1 \hdots a_{n-1} ) 
- \mathbf{V}( \theta_{n-1} | a_1 \hdots a_{n-1} )
\end{align}
after the $n$'th interrogation. 

For each interrogation, we obtain the optimal quantum algorithm by
extending the procedure described in Ref.~\cite{firstClock}.  There,
we relate the operation of an atomic clock to quantum complexity
theory, specifically a generalization of the adversary method, and use
this relationship to calculate quantum algorithms that optimize the
expected posterior cost of an interrogation,
\begin{equation} 
 \mathbf{E}(C) =
  \sum_a \int C(\omega, a) p(\omega,a) d\omega.
\label{eq:clockCost} 
\end{equation}
Here, $\omega$ is the frequency deviation of the external oscillator
during the interrogation of interest. While we can optimize the cost
for any reasonable cost function, the choice is determined by how we
quantify clock performance.  A traditional choice and the one
emphasized in Ref.~\cite{firstClock} is $C(\omega, a) = ( \omega -
g(a) )^2$, where the $g(a)$ are frequency estimates depending on the
(arbitrarily labeled) measurement outcomes $a$.  The estimates can be
chosen so that minimization of $\mathbf{E}(C)$ for the $n$'th
interrogation minimizes the expected posterior variance of $\omega_n
T_n$.  However, for noise models with memory, this does not minimize
the expected posterior total variance increase $\Delta V_n$. This is
because in general, $\mathbf{V}( \theta_n | \boldsymbol{a_n} )$ is not
the same as $\mathbf{V}( \omega_1T_1 | a_1 ) + \mathbf{V}( \omega_2T_2
| a_2 ) + \hdots \mathbf{V}( \omega_nT_n | a_n )$, due to correlations
between the $\omega_i$'s.  In App.~\ref{app:A}, we
prove that the following adaptively chosen cost function has the
desired effect of minimizing $\Delta V_n$:
\begin{equation}\label{eq:costFunction}
C(\omega, a) = (\omega T - g(a) )^2 + 2 ( \omega T - g(a) )\mathbf{E}( \theta - \mathbf{E}(\theta) | \omega ),
\end{equation}
where $\theta$ is the phase deviation just before the interrogation of
interest and $\omega$ is the frequency of the oscillator during this
interrogation. The expectations in the cost function are implicitly
conditioned on every earlier measurement outcome. The minimum $\Delta
V_{n}$ is achieved in the continuum limit of the SDP, where the
measurement outcome labels are possible average frequencies
$\omega_{n}$ and $g$ is the identity function. The implemented SDPs
involve discretization. Ref.~\cite{firstClock} shows that the
discretization error can be made arbitrarily small and how to bound
it.

In order to derive algorithms that minimize $\Delta V_{n}$, we need
access to $p( \boldsymbol{\omega_n} | \boldsymbol{a_{n-1}})$ before
the $n$'th interrogation.  This requires that we correctly maintain
and update such a distribution as a clock runs.  For the moment, we
assume that it is possible to keep track of these continous and
high-dimensional distributions exactly. Later, we discuss how to
discretize and truncate them in practice.  Fig.~\ref{fig:flowchart}
depicts the evolution of this probability distribution associated with
the first interrogation.  In general, before the $n$'th interrogation,
we have access to $p( \boldsymbol{ \omega_{n-1}} | \boldsymbol{
  a_{n-1} } )$ as computed from the previous interrogation or, for
$n=1$, from the initial conditions.  The $n$'th interrogation requires
that we (1) compute the prior $p( \boldsymbol{ \omega_n } |
\boldsymbol{ a_{n-1} } )$ according to the noise model and previously
determined priors and measurement outcomes, (2) derive and apply a
quantum algorithm based on this distribution, and (3) compute the
posterior distribution $p( \boldsymbol{ \omega_n} | \boldsymbol{ a_n }
)$ from the prior and measurement outcome $a_n$. In more detail,
the procedure is:

\begin{figure}[ht]
\includegraphics[width=.75\textwidth]{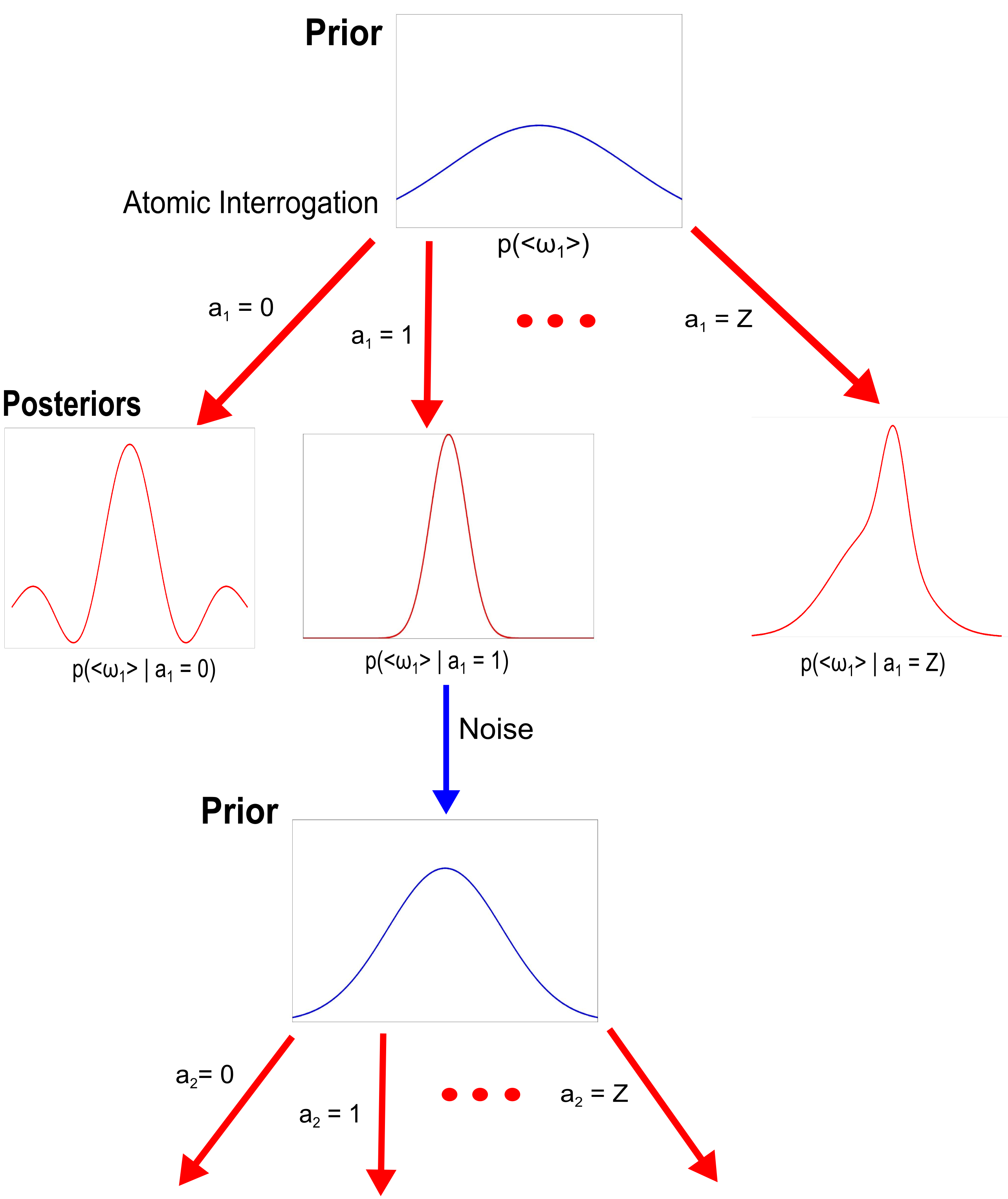}
\caption{Evolution of the external oscillator's posterior probability
  distribution in a clock protocol. The probability distribution at
  the top characterizes the frequency of the clock during
  interrogation $1$.  A measurement of the atomic standard at time
  $t_{1}$, with $Z$ element POVM $\{P_a\}_a$ then yields measurement
  outcome $a_1$ with probability $p(a_1)$.  This can be used to
  compute the posterior distribution $p( \omega_1 |a_1)$.  (In the
  figure, we imagine that we obtain measurement outcome $a_1 = 1$.)
  Noise then affects the clock for duration $T_2$; this decreases our
  knowledge of the external oscillator's frequency, widening the
  probability distribution.  The resulting prior probability
  distribution corresponds to the frequency of the classical
  oscillator during interrogation $2$.  This procedure is repeated
  indefinitely. }
\label{fig:flowchart}
\end{figure}

\begin{enumerate}
\item Compute the prior probability distribution $p(
  \boldsymbol{\omega_{n}} | \boldsymbol{a_{n-1}} )$
  according to 
  \begin{equation}
    p(\boldsymbol{\omega_{n}} | \boldsymbol{a_{n-1}} )= p( \omega_{n} |
    \boldsymbol{\omega_{n-1}}, \boldsymbol{a_{n-1}} ) p(
    \boldsymbol{\omega_{n-1}} | \boldsymbol{a_{n-1}} ).
  \end{equation}
  For this purpose, note that $p( \omega_{n} |
  \boldsymbol{\omega_{n-1}}, \boldsymbol{a_{n-1}} ) = p( \omega_{n} |
  \boldsymbol{\omega_{n-1}})$, so that it can be computed directly
  from the noise model's covariance matrix.

\item From $p( \boldsymbol{\omega_{n}} | \boldsymbol{a_{n-1}})$,
  determine the cost function of Eq.~\eqref{eq:costFunction} and apply
  the SDP of Ref. \cite{firstClock} to compute an optimal algorithm
  consisting of a unitary matrix $U$ and a POVM $\{P_a\}_{a}$.

\item Given this algorithm, fill in the collection of
  distributions $p(a_{n} | \omega_{n} )$ for each $\omega_{n}$.

\item Use the algorithm to interrogate the frequency standard for a
  time $T_n$.  Obtain the actual measurement outcome $a^*_n$.

\item Assign $a_n=a^*_n$ and compute the posterior distribution needed
  for the next timestep according to
  \begin{align}
    p( \boldsymbol{\omega_{n}} | \boldsymbol{a_{n}}) &= p(\omega_{n},
    \boldsymbol{\omega_{n-1}} | a_n, \boldsymbol{a_{n-1}} )\nonumber\\
    &=p(\omega_{n},
    \boldsymbol{\omega_{n-1}}, a_{n}| \boldsymbol{a_{n-1}} )/
    p(a_{n}|\boldsymbol{a_{n-1}})\nonumber\\
    &= p(\omega_{n},\boldsymbol{\omega_{n-1}}| \boldsymbol{a_{n-1}} )
    p( a_{n} | \omega_{n}, \boldsymbol{\omega_{n-1}},
    \boldsymbol{a_{n-1}} )/p(a_{n}|\boldsymbol{a_{n-1}})\nonumber\\
    &= p(\boldsymbol{\omega_{n}}|\boldsymbol{a_{n-1}})
       p(a_{n}|\omega_{n})/p(a_{n}|\boldsymbol{a_{n-1}}).
  \end{align}
  Here, we used the fact that $a_{n}$ is independent of
  $\boldsymbol{\omega_{n-1}}$ and $\boldsymbol{a_{n-1}}$ given
  $\omega_{n}$.  The term $p(a_{n}|\boldsymbol{a_{n-1}})$ can
  be computed as
  \begin{align}
    p(a_{n}|\boldsymbol{a_{n-1}}) &=\int d{\boldsymbol{\omega'_{n}}}
    p(a_{n}|\boldsymbol{\omega'_{n}},\boldsymbol{a_{n-1}})
    p(\boldsymbol{\omega'_{n}}|\boldsymbol{a_{n-1}})\nonumber\\
    &=\int d{\boldsymbol{\omega'_{n}}}
    p(a_{n}|\omega'_{n}) p(\boldsymbol{\omega'_{n}}|\boldsymbol{a_{n-1}}).
  \end{align}

\item Compute the posterior expectation $\mathbf{E}(\theta_n |
  \boldsymbol{a_{n}} )$ of the phase $\theta_{n} = \theta_{n-1} +
  \omega_n T_n$.  This is our estimate of the external oscillator's
  phase after interrogation $n$ and may be used to assign timestamps.
\end{enumerate}

Note that this procedure can be readily generalized if other
information becomes available during an interrogation. Here, the part
of the clock's state relevant to timekeeping given the interrogation
history is determined by the (true) frequencies $\omega_{n}$. In
general, there may be other state variables we can exploit, in which
case the relevant part of the state is given by more fundamental
variables $s_{n}$ describing the state during the $n$'th
interrogation. Also, after each interrogation, the best estimates of
the phases $\theta_{k}$ for $k<n$ based on current information can
change. Thus, it is beneficial to retroactively update these estimates
also.

\section{Systematic Errors}

There are three sources of error that arise in the above interrogation
procedure: (1) Discretization error in the SDP used to construct each
unitary and POVM, (2) discretization and truncation of
$p(\boldsymbol{\omega_n} | \boldsymbol{a_n} )$, and (3) incomplete
knowledge of the true duration of each interrogation.  The first issue
was discussed in Ref.~\cite{firstClock}; here, we discuss the other
two.

To address the second source of error, note that the distributions of
the $\omega_n$ are inherently continuous and must be discretized
sufficiently finely.  However, if we discretize the domain of
$\omega_n$ with $P$ points, then the representation of the joint
probability distribution grows by a factor of $P$ at every step; if
$P$ is large, the strategy described above quickly becomes
computationally infeasible.  We therefore truncate the clock history
by storing only a limited number of $\omega_n$, and marginalizing out
old distributions as the clock progresses.  Since most of the noise
models discussed above contain long-term correlations, this procedure
no longer represents these models faithfully.  But the correlations
typically fall off as a power law, so we may be justified in
concluding that the impact on the performance of our protocol is
limited, provided enough memory is maintained. The truncation
indirectly affects the SDP. While the SDP does not explicitly require
the full joint distribution, the cost function of
Eq.~\eqref{eq:costFunction} involves expectations of the cumulative
phase $\theta_{n-1}$ and depends on the $\omega_{k}$ lost in
truncation.  In App.~\ref{app:C} we show how this expectation, and more
generally, $\mathbf{E}(\theta_n^k | \omega_{n+1}, \boldsymbol{a_n})$
for arbitrary $k$ can be updated without keeping full track of all
$\omega_{k}$.

With regard to the third issue, so far we have fixed the duration of
interrogation $n$ at $T_n$ and assumed that $T_{n}$ is the ``real''
duration.  However, the end points of the interrogation are
chosen by the experimenter based on the external oscillator or an
auxiliary clock locked to the oscillator.  In addition, the
implementations of state preparation and measurement take finite time,
adding additional uncertainty concerning the true duration of the
implemented interrogation. The standard interrogation methods are
normally insensitive to variations in $T_{n}$ and non-zero preparation
and measurement intervals because the external oscillator's frequency is
constantly controlled to match the frequency standard.  For our
protocols, explicitly changing the external oscillator's frequency
within the memory time of the noise model would complicate the
algorithm for keeping track of the relevant posterior probability
distributions. With a free-running external oscillator, it is
necessary to adapt the interrogation algorithms to minimize the effect
of timing deviations. One adaptation involves simulating the effect of
a locked oscillator. We also suggest that it is beneficial to adapt
the SDP used to optimize the interrogations. How to implement both
adaptations and the size of residual errors is discussed in the 
App.~\ref{app:D}.

\section{Simulations}

To test our protocol, we implemented a general-purpose Monte Carlo
simulation of the external oscillator and used it in a simulated clock
with the above protocol and update strategy. In these simulations, we
used a constant interrogation duration $T$ throughout.  To evaluate the
simulated clocks, we compute the average square difference between the
estimated average frequency and the true average frequency of the
simulated external oscillator (the ``square frequency error''),
where both are cumulative time-averages from the start of the clock. We also compute the overlapping Allan variances given by
\begin{align}
\label{eq:overlappedAllanVarianceF}
\sigma^2(mT) =& \frac{1}{2(M - 2m + 1)}\times\nonumber\\
&\sum_{j=1}^{M-2m+1}(\langle \omega^{*}\rangle_{(j+m,m)} - \langle \omega^{*} \rangle_{(j,m)} )^2,
\end{align}
where $M$ is the total number of interrogations, each of equal
duration $T$, $\langle
\omega^{*}\rangle_{k,m}=\sum_{l=k}^{k+m-1}\omega^{*}_{l}/m$ and
$\omega^{*}_{j}$ is the best estimate of $\omega_{j}$ given by the
computed mean of the relevant posterior probability distributions.
The Allan variance is what would actually be reported in an
experimental realization of these clocks and does not depend on
knowing the true frequencies.

\begin{figure*}
\resizebox{1\textwidth}{!}{
\subfigure[\ Root-Square Frequency Error ]{
\includegraphics[width=.4\textwidth]{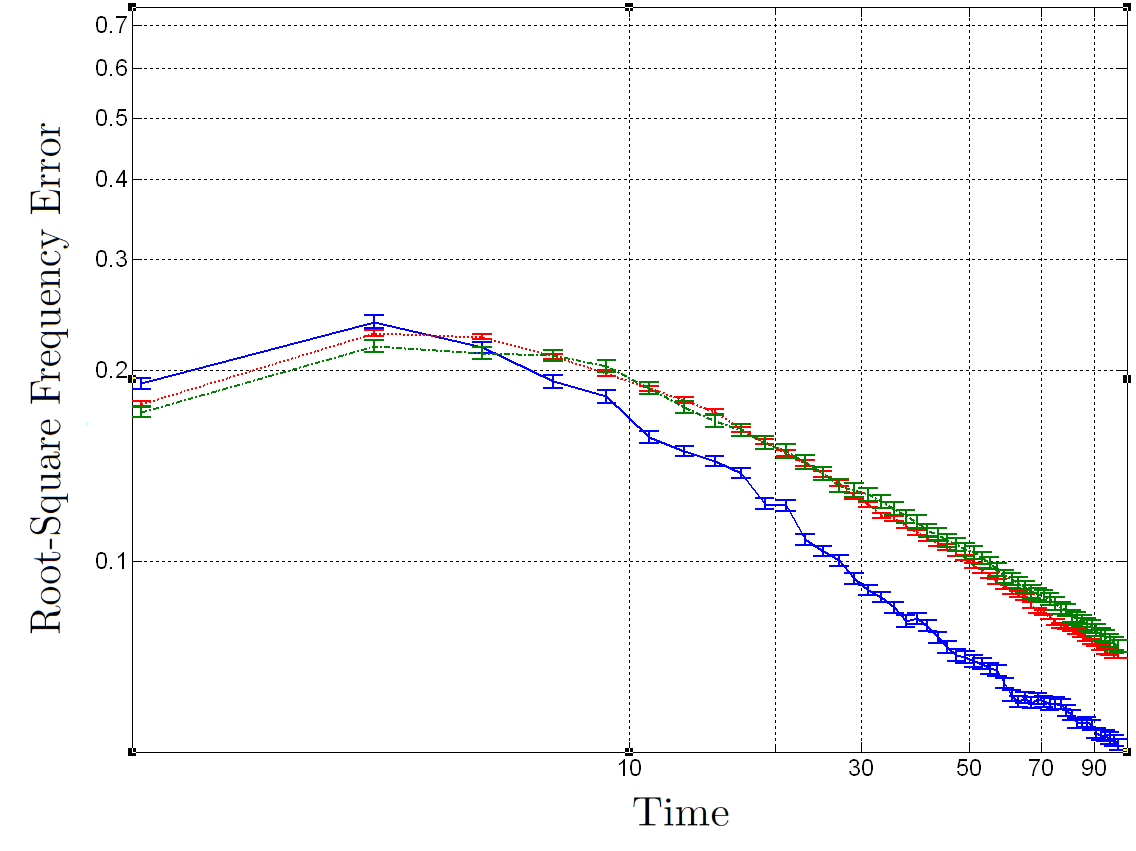}}
\subfigure[\  Overlapping Allan Deviation ]{
\includegraphics[width=.4\textwidth]{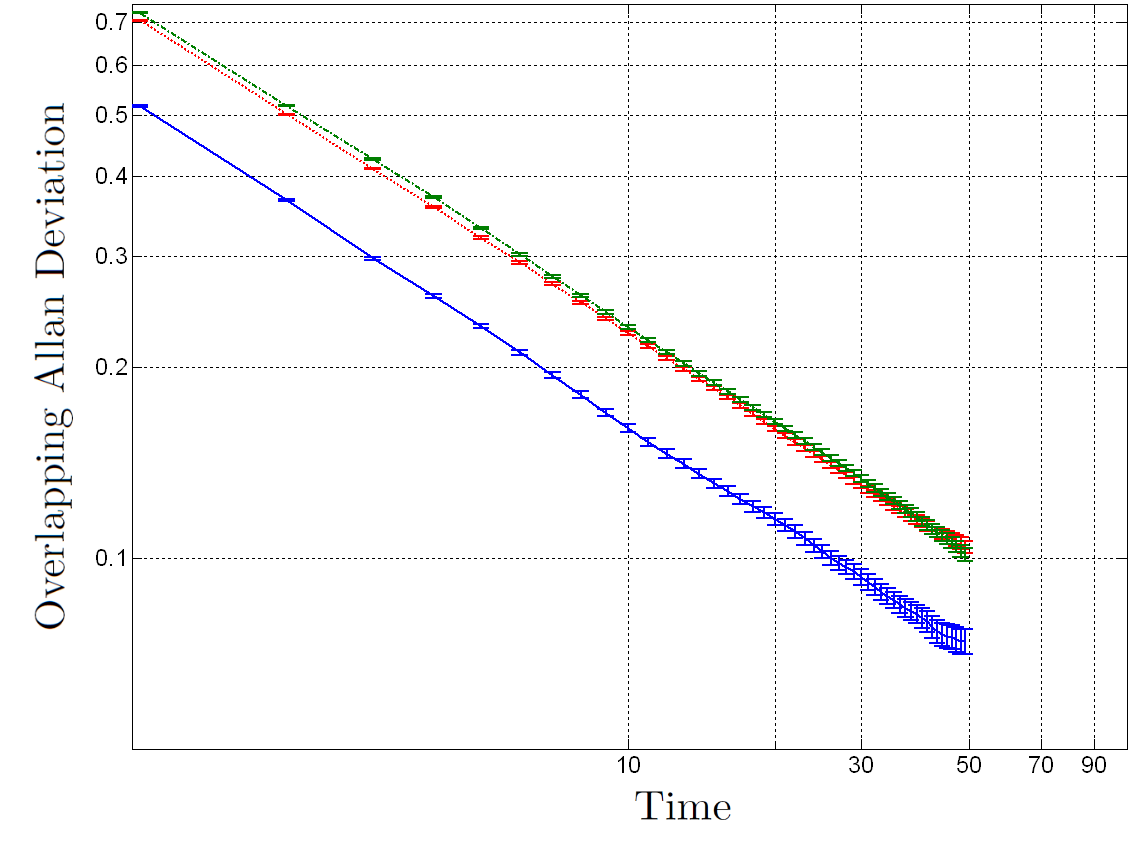}}
}
\caption{A comparison of our protocol (bottom blue solid line), the
  standard Ramsey protocol (middle red dotted line), and Buzek's
  protocol~\cite{buzek} (top green dashed line) on a two atom clock
  subject to standard Brownian motion with $h_{-2} = .03$. We fix the
  length of the interrogation to $T=1$.  Figure (a) is a log-log plot
  of the root-square error of the frequency averaged over
  cumulative time, while (b) is a log-log plot of the overlapping
  Allan deviation with respect to $m$.  These are computed with $200$
  runs of our method and $1000$ of each of Ramsey's and Buzek's
  protocol.  }
\label{fig:brownian}
\end{figure*}

Below (see Fig.~\ref{fig:brownian}), we compare our protocol to the
Ramsey protocol, which is utilized by most atomic clocks today, and to
that of Buzek et. al.~\cite{buzek}.  The latter is a fully quantum
technique optimized for a uniform prior probability distribution of
external oscillator frequencies. We limit our comparisons to clocks
with low noise in order to reduce phase-slip errors that result in
random frequency hops of size $2\pi/T$.

The traditional Ramsey protocol is used with an external oscillator
that is controlled to have a frequency matching the atomic standard as
closely as possible. To simplify noise model calculations, we do not
adjust the external oscillator. Instead we compute the measurement
phase directly, according to the computed means of the prior
probability distribution for the frequencies. See the discussion of
timing errors in App.~\ref{app:D}.  Provided the noise model is a good
representation of the external oscillator's behavior, this is expected
to perform better than the standard control strategies, so that our
comparison is fair.

\begin{figure*}
\resizebox{1\textwidth}{!}{
\subfigure[\ Root-Square Frequency Error ]{
\includegraphics[width=.4\textwidth]{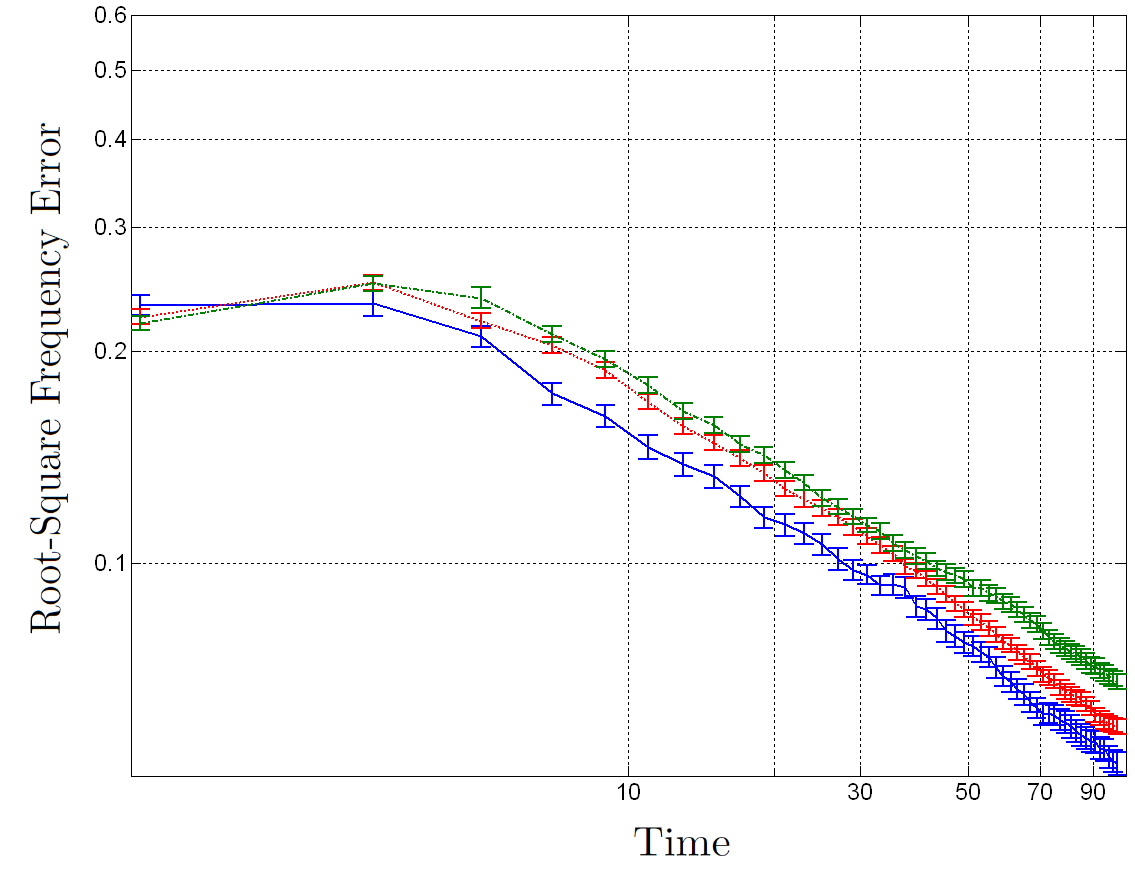}}
\subfigure[\   Overlapping Allan Deviation ]{
\includegraphics[width=.4\textwidth]{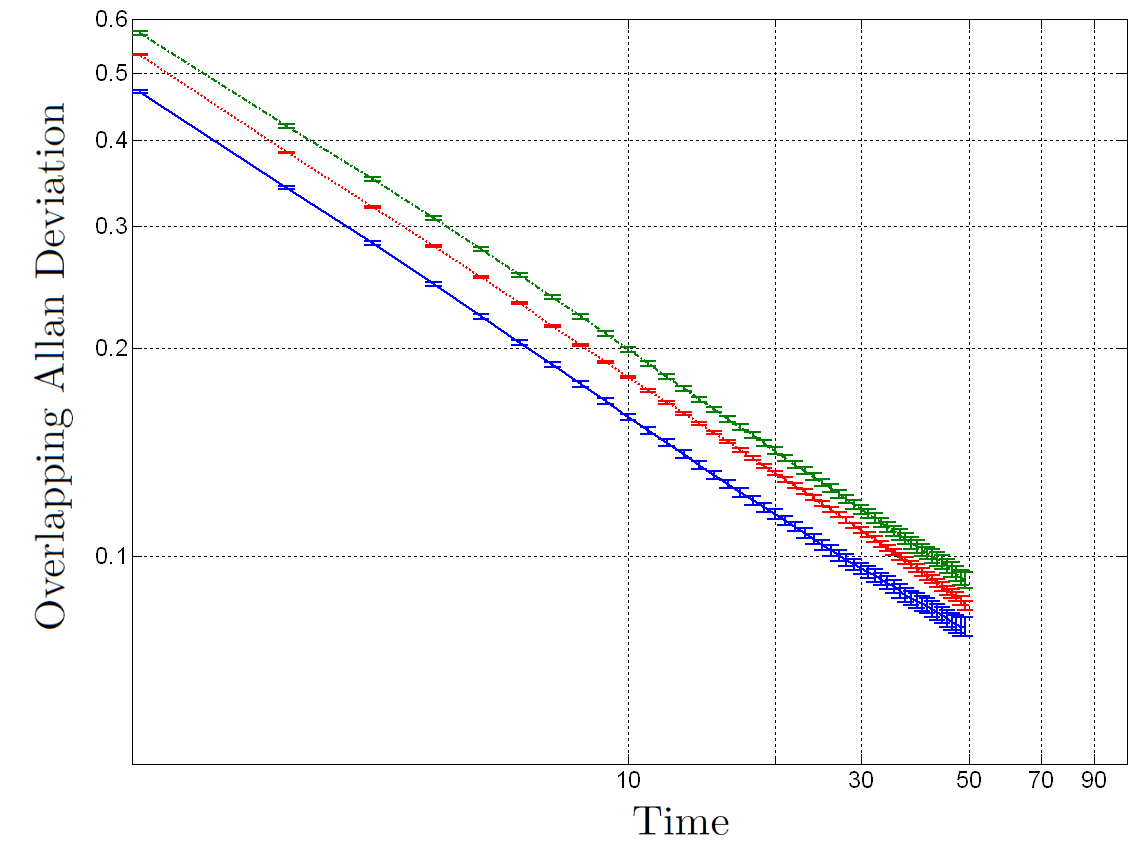}}
}
\caption{A comparison of our protocol (bottom blue solid line), the
  standard Ramsey protocol (middle red dotted line), and Buzek's
  protocol (top green dashed line) on a three atom clock subject to
  $1/\omega$ noise with $h_{-1} = .05$.  Plots (a) and (b) are as
  described in Fig. \ref{fig:brownian} and are computed via $400$
  iterations of our protocol and $800$ iterations of each of Ramsey's
  and Buzek's protocols. 
}
\label{fig:1f}
\end{figure*}

Fig.~\ref{fig:brownian} compares our protocol to that of Ramsey and
that of Buzek for a two-atom clock subject to Brownian motion ($\alpha
= -2$) with $h_{-2}=.03$ and $100$ interrogations.  Since Brownian
motion is memoryless, keeping a history of just the last interrogation
suffices.  Fig.~\ref{fig:1f} shows the comparison for a three-atom
clock subject to $1/\omega$ noise, ($\alpha = -1$), with $h_{-1}=.05$.
We cannot maintain the infinite history required by this noise model
and truncate the frequency history after one step.  Note that we
expect Buzek's protocol to perform significantly better in clocks with
large numbers of atoms.  Table~\ref{table:gain} summarizes the
improvements achieved by our technique.  These results are consistent
with those predicted in Ref.~\cite{numericalTest}. We expect greater
improvements by storing a more complete frequency history, by using
multi-round strategies, and in clocks with additional atoms.

\begin{table}[h]
\centering
\begin{tabular} {c | c | c c}
\textbf{Noise Type} & & \textbf{Ramsey} & \textbf{Buzek} \\
\hline
Brownian & Square Error & $46.9 \pm .51$ & $51.7 \pm .46$ \\ \hline
Brownian & Allan Variance &  $49.3 \pm .43$ & $50.9 \pm .42$ \\ \hline
$1/f$ & Square Error & $18.9 \pm .40$ & $39.8 \pm .38$ \\ \hline
$1/f$ & Allan Variance & $21.6 \pm .55$ & $33.0 \pm .48$ 
\end{tabular}
\caption{
  Percent improvement of our protocol over those of Ramsey and Buzek.
  We average the gains in square error over the last twenty
  timesteps and those in Allan variance across all averaging times.
}
\label{table:gain}
\end{table}

\section{Conclusion}

While the protocols discussed here already significantly outperform
traditional clock protocols, we can obtain further improvements by
choosing the interrogation duration $T$ optimally at every step.  Longer
interrogation durations can provide more information, but if $T$ is chosen
too large the clock's frequency can slip.  While this issue is
beyond the scope of this paper, we believe our protocol can be adjusted
to choose $T$ adaptively. Also, while our protocol performs well
even when used with a significantly truncated frequency history,
additional storage would, nonetheless, be advantageous. Unfortunately,
this often requires a dramatic increase in computation time.  It will
be helpful to investigate this tradeoff in more detail, and ideally,
develop a systematic way to determine when to cut off the clock's
history.

The protocols we have developed were implemented on simulated clocks
as a proof-of-principle. Application to an experimental setting
requires that the interrogation algorithms obtained be converted to
the elementary quantum control operations actually available.  Since
the interrogation algorithms are different for each timestep, they
need to be converted to atom-control operations on the fly.  The
conversion should optimize control-related decoherence, accuracy of
the implemented evolutions, and time.  This seems feasible for small
numbers of atoms in a sufficiently controllable setting. For more
atoms, the optimal interrogation algorithms obtained may be too
complex to be implemented with sufficiently low error. It will be
necessary to optimize the interrogations in view of limited
experimental resources. In practice, it is possible that most of the
gains achieved by the protocols can be realized with a restricted
set of pre-optimized interrogation algorithms. The effects of the
necessary compromises on clock performance need to be investigated.

We conclude by noting that some of the most accurate clocks now being
developed use a small number of ions~\cite{clockOverview,
  PhysRevLett.104.070802}.  Full quantum control over systems of
comparable size has already been demonstrated in ion
traps~\cite{2QubitIon}.  We therefore expect quantum techniques to be
experimentally applicable relatively soon.  This is in contrast to
other domains in which quantum algorithms have been theoretically
shown to offer advantages, but where solving useful instances of
interesting problems requires control over quantum systems of sizes
far beyond what is currently achievable experimentally.  Indeed, clocks may be
among the first systems where a nontrivial quantum algorithmic gain is
realized.

\bibliographystyle{apsrev}
\bibliography{clockSimulations}

\begin{thebibliography}{28}
\expandafter\ifx\csname natexlab\endcsname\relax\def\natexlab#1{#1}\fi
\expandafter\ifx\csname bibnamefont\endcsname\relax
  \def\bibnamefont#1{#1}\fi
\expandafter\ifx\csname bibfnamefont\endcsname\relax
  \def\bibfnamefont#1{#1}\fi
\expandafter\ifx\csname citenamefont\endcsname\relax
  \def\citenamefont#1{#1}\fi
\expandafter\ifx\csname url\endcsname\relax
  \def\url#1{\texttt{#1}}\fi
\expandafter\ifx\csname urlprefix\endcsname\relax\def\urlprefix{URL }\fi
\providecommand{\bibinfo}[2]{#2}
\providecommand{\eprint}[2][]{\url{#2}}

\bibitem[{\citenamefont{Wineland et~al.}(1992)\citenamefont{Wineland,
  Bollinger, Itano, Moore, and Heinzen}}]{squeeze3}
\bibinfo{author}{\bibfnamefont{D.}~\bibnamefont{Wineland}},
  \bibinfo{author}{\bibfnamefont{J.}~\bibnamefont{Bollinger}},
  \bibinfo{author}{\bibfnamefont{W.}~\bibnamefont{Itano}},
  \bibinfo{author}{\bibfnamefont{F.}~\bibnamefont{Moore}}, \bibnamefont{and}
  \bibinfo{author}{\bibfnamefont{D.}~\bibnamefont{Heinzen}},
  \bibinfo{journal}{Physical Review A} \textbf{\bibinfo{volume}{46}},
  \bibinfo{pages}{6797} (\bibinfo{year}{1992}).

\bibitem[{\citenamefont{Kitagawa and Ueda}(1993)}]{squeeze}
\bibinfo{author}{\bibfnamefont{M.}~\bibnamefont{Kitagawa}} \bibnamefont{and}
  \bibinfo{author}{\bibfnamefont{M.}~\bibnamefont{Ueda}},
  \bibinfo{journal}{Physical review. A} \textbf{\bibinfo{volume}{47}},
  \bibinfo{pages}{5138} (\bibinfo{year}{1993}).

\bibitem[{\citenamefont{Wineland et~al.}(1994)\citenamefont{Wineland,
  Bollinger, Itano, and Heinzen}}]{squeeze2}
\bibinfo{author}{\bibfnamefont{D.}~\bibnamefont{Wineland}},
  \bibinfo{author}{\bibfnamefont{J.}~\bibnamefont{Bollinger}},
  \bibinfo{author}{\bibfnamefont{W.}~\bibnamefont{Itano}}, \bibnamefont{and}
  \bibinfo{author}{\bibfnamefont{D.}~\bibnamefont{Heinzen}},
  \bibinfo{journal}{Physical Review A} \textbf{\bibinfo{volume}{50}},
  \bibinfo{pages}{67} (\bibinfo{year}{1994}).

\bibitem[{\citenamefont{Bollinger et~al.}(1996)\citenamefont{Bollinger, Itano,
  Wineland, and Heinzen}}]{bollinger:qc1996a}
\bibinfo{author}{\bibfnamefont{J.~J.} \bibnamefont{Bollinger}},
  \bibinfo{author}{\bibfnamefont{W.~M.} \bibnamefont{Itano}},
  \bibinfo{author}{\bibfnamefont{D.~J.} \bibnamefont{Wineland}},
  \bibnamefont{and} \bibinfo{author}{\bibfnamefont{D.~J.}
  \bibnamefont{Heinzen}}, \bibinfo{journal}{Phys. Rev. A}
  \textbf{\bibinfo{volume}{54}}, \bibinfo{pages}{R4649} (\bibinfo{year}{1996}).

\bibitem[{\citenamefont{Huelga et~al.}(1997)\citenamefont{Huelga, Macchiavello,
  Pellizzari, Ekert, Plenio, and Cirac}}]{huelga}
\bibinfo{author}{\bibfnamefont{S.}~\bibnamefont{Huelga}},
  \bibinfo{author}{\bibfnamefont{C.}~\bibnamefont{Macchiavello}},
  \bibinfo{author}{\bibfnamefont{T.}~\bibnamefont{Pellizzari}},
  \bibinfo{author}{\bibfnamefont{A.}~\bibnamefont{Ekert}},
  \bibinfo{author}{\bibfnamefont{M.}~\bibnamefont{Plenio}}, \bibnamefont{and}
  \bibinfo{author}{\bibfnamefont{J.}~\bibnamefont{Cirac}},
  \bibinfo{journal}{Phys. Rev. Lett.} \textbf{\bibinfo{volume}{79}},
  \bibinfo{pages}{3865} (\bibinfo{year}{1997}).

\bibitem[{\citenamefont{Berry et~al.}(2009)\citenamefont{Berry, Higgins,
  Bartlett, Mitchell, Pryde, and Wiseman}}]{noisyProtocol}
\bibinfo{author}{\bibfnamefont{D.}~\bibnamefont{Berry}},
  \bibinfo{author}{\bibfnamefont{B.}~\bibnamefont{Higgins}},
  \bibinfo{author}{\bibfnamefont{S.}~\bibnamefont{Bartlett}},
  \bibinfo{author}{\bibfnamefont{M.}~\bibnamefont{Mitchell}},
  \bibinfo{author}{\bibfnamefont{G.}~\bibnamefont{Pryde}}, \bibnamefont{and}
  \bibinfo{author}{\bibfnamefont{H.}~\bibnamefont{Wiseman}},
  \bibinfo{journal}{Physical Review A} \textbf{\bibinfo{volume}{80}},
  \bibinfo{pages}{052114} (\bibinfo{year}{2009}).

\bibitem[{\citenamefont{Higgins et~al.}(2007)\citenamefont{Higgins, Berry,
  Bartlett, Wiseman, and Pryde}}]{noisyProtocol2}
\bibinfo{author}{\bibfnamefont{B.}~\bibnamefont{Higgins}},
  \bibinfo{author}{\bibfnamefont{D.}~\bibnamefont{Berry}},
  \bibinfo{author}{\bibfnamefont{S.}~\bibnamefont{Bartlett}},
  \bibinfo{author}{\bibfnamefont{H.}~\bibnamefont{Wiseman}}, \bibnamefont{and}
  \bibinfo{author}{\bibfnamefont{G.}~\bibnamefont{Pryde}},
  \bibinfo{journal}{Nature} \textbf{\bibinfo{volume}{450}},
  \bibinfo{pages}{393} (\bibinfo{year}{2007}).

\bibitem[{\citenamefont{Huver et~al.}(2008)\citenamefont{Huver, Wildfeuer, and
  Dowling}}]{noisyProtocol3}
\bibinfo{author}{\bibfnamefont{S.~D.} \bibnamefont{Huver}},
  \bibinfo{author}{\bibfnamefont{C.~F.} \bibnamefont{Wildfeuer}},
  \bibnamefont{and} \bibinfo{author}{\bibfnamefont{J.~P.}
  \bibnamefont{Dowling}}, \bibinfo{journal}{Phys. Rev. A}
  \textbf{\bibinfo{volume}{78}}, \bibinfo{pages}{063828}
  (\bibinfo{year}{2008}),
  \urlprefix\url{http://link.aps.org/doi/10.1103/PhysRevA.78.063828}.

\bibitem[{\citenamefont{Luis}(2002)}]{noisyProtocol4}
\bibinfo{author}{\bibfnamefont{A.}~\bibnamefont{Luis}}, \bibinfo{journal}{Phys.
  Rev. A} \textbf{\bibinfo{volume}{65}}, \bibinfo{pages}{025802}
  (\bibinfo{year}{2002}),
  \urlprefix\url{http://link.aps.org/doi/10.1103/PhysRevA.65.025802}.

\bibitem[{\citenamefont{Dorner}(2012)}]{noisyProtocol5}
\bibinfo{author}{\bibfnamefont{U.}~\bibnamefont{Dorner}}, \bibinfo{journal}{New
  Journal of Physics} \textbf{\bibinfo{volume}{14}}, \bibinfo{pages}{043011}
  (\bibinfo{year}{2012}).

\bibitem[{\citenamefont{Dorner et~al.}(2009)\citenamefont{Dorner,
  Demkowicz-Dobrzanski, Smith, Lundeen, Wasilewski, Banaszek, and
  Walmsley}}]{noisyProtocol6}
\bibinfo{author}{\bibfnamefont{U.}~\bibnamefont{Dorner}},
  \bibinfo{author}{\bibfnamefont{R.}~\bibnamefont{Demkowicz-Dobrzanski}},
  \bibinfo{author}{\bibfnamefont{B.~J.} \bibnamefont{Smith}},
  \bibinfo{author}{\bibfnamefont{J.~S.} \bibnamefont{Lundeen}},
  \bibinfo{author}{\bibfnamefont{W.}~\bibnamefont{Wasilewski}},
  \bibinfo{author}{\bibfnamefont{K.}~\bibnamefont{Banaszek}}, \bibnamefont{and}
  \bibinfo{author}{\bibfnamefont{I.~A.} \bibnamefont{Walmsley}},
  \bibinfo{journal}{Phys. Rev. Lett.} \textbf{\bibinfo{volume}{102}},
  \bibinfo{pages}{040403} (\bibinfo{year}{2009}),
  \urlprefix\url{http://link.aps.org/doi/10.1103/PhysRevLett.102.040403}.

\bibitem[{\citenamefont{Wineland et~al.}(1998)\citenamefont{Wineland, Monroe,
  Itano, Leibfried, King, and Meekhof}}]{winelandBible}
\bibinfo{author}{\bibfnamefont{D.~J.} \bibnamefont{Wineland}},
  \bibinfo{author}{\bibfnamefont{C.}~\bibnamefont{Monroe}},
  \bibinfo{author}{\bibfnamefont{W.~M.} \bibnamefont{Itano}},
  \bibinfo{author}{\bibfnamefont{D.}~\bibnamefont{Leibfried}},
  \bibinfo{author}{\bibfnamefont{B.~E.} \bibnamefont{King}}, \bibnamefont{and}
  \bibinfo{author}{\bibfnamefont{D.~M.} \bibnamefont{Meekhof}},
  \bibinfo{journal}{J. Res. NIST} \textbf{\bibinfo{volume}{103}},
  \bibinfo{pages}{259} (\bibinfo{year}{1998}).

\bibitem[{\citenamefont{Andr\'e et~al.}(2004)\citenamefont{Andr\'e,
  S\o{}rensen, and Lukin}}]{squeezedClocks}
\bibinfo{author}{\bibfnamefont{A.}~\bibnamefont{Andr\'e}},
  \bibinfo{author}{\bibfnamefont{A.~S.} \bibnamefont{S\o{}rensen}},
  \bibnamefont{and} \bibinfo{author}{\bibfnamefont{M.~D.} \bibnamefont{Lukin}},
  \bibinfo{journal}{Phys. Rev. Lett.} \textbf{\bibinfo{volume}{92}},
  \bibinfo{pages}{230801} (\bibinfo{year}{2004}).

\bibitem[{\citenamefont{{Bu{\v z}ek} et~al.}(1999)\citenamefont{{Bu{\v z}ek},
  {Derka}, and {Massar}}}]{buzek}
\bibinfo{author}{\bibfnamefont{V.}~\bibnamefont{{Bu{\v z}ek}}},
  \bibinfo{author}{\bibfnamefont{R.}~\bibnamefont{{Derka}}}, \bibnamefont{and}
  \bibinfo{author}{\bibfnamefont{S.}~\bibnamefont{{Massar}}},
  \bibinfo{journal}{Phys. Rev. Lett.} \textbf{\bibinfo{volume}{82}},
  \bibinfo{pages}{2207} (\bibinfo{year}{1999}),
  \eprint{arXiv:quant-ph/9808042}.

\bibitem[{\citenamefont{Demkowicz-Dobrza\ifmmode~\acute{n}\else
  \'{n}\fi{}ski}(2011)}]{dobrzanski}
\bibinfo{author}{\bibfnamefont{R.}~\bibnamefont{Demkowicz-Dobrza\ifmmode~\acute{n}\else
  \'{n}\fi{}ski}}, \bibinfo{journal}{Phys. Rev. A}
  \textbf{\bibinfo{volume}{83}}, \bibinfo{pages}{061802}
  (\bibinfo{year}{2011}),
  \urlprefix\url{http://link.aps.org/doi/10.1103/PhysRevA.83.061802}.

\bibitem[{\citenamefont{van Dam et~al.}(2007)\citenamefont{van Dam, D'Ariano,
  Ekert, Macchiavello, and Mosca}}]{van2007optimal}
\bibinfo{author}{\bibfnamefont{W.}~\bibnamefont{van Dam}},
  \bibinfo{author}{\bibfnamefont{G.}~\bibnamefont{D'Ariano}},
  \bibinfo{author}{\bibfnamefont{A.}~\bibnamefont{Ekert}},
  \bibinfo{author}{\bibfnamefont{C.}~\bibnamefont{Macchiavello}},
  \bibnamefont{and} \bibinfo{author}{\bibfnamefont{M.}~\bibnamefont{Mosca}},
  \bibinfo{journal}{Phy. Rev. Lett.} \textbf{\bibinfo{volume}{98}},
  \bibinfo{pages}{90501} (\bibinfo{year}{2007}).

\bibitem[{\citenamefont{Macieszczak et~al.}(2013)\citenamefont{Macieszczak,
  Demkowicz-Dobrzanski, and Fraas}}]{macieszczak2013optimal}
\bibinfo{author}{\bibfnamefont{K.}~\bibnamefont{Macieszczak}},
  \bibinfo{author}{\bibfnamefont{R.}~\bibnamefont{Demkowicz-Dobrzanski}},
  \bibnamefont{and} \bibinfo{author}{\bibfnamefont{M.}~\bibnamefont{Fraas}},
  \bibinfo{journal}{arXiv preprint arXiv:1311.5576}  (\bibinfo{year}{2013}).

\bibitem[{\citenamefont{Rosenband}(2012)}]{numericalTest}
\bibinfo{author}{\bibfnamefont{T.}~\bibnamefont{Rosenband}}
  (\bibinfo{year}{2012}), \bibinfo{note}{arXiv:1203.0288v2}.

\bibitem[{\citenamefont{Giovannetti et~al.}(2011)\citenamefont{Giovannetti,
  Lloyd, and Maccone}}]{giovannetti:qc2011a}
\bibinfo{author}{\bibfnamefont{V.}~\bibnamefont{Giovannetti}},
  \bibinfo{author}{\bibfnamefont{S.}~\bibnamefont{Lloyd}}, \bibnamefont{and}
  \bibinfo{author}{\bibfnamefont{L.}~\bibnamefont{Maccone}},
  \bibinfo{journal}{Nat. Phot.} \textbf{\bibinfo{volume}{5}},
  \bibinfo{pages}{222} (\bibinfo{year}{2011}).

\bibitem[{\citenamefont{Mullan and Knill}(2012)}]{firstClock}
\bibinfo{author}{\bibfnamefont{M.}~\bibnamefont{Mullan}} \bibnamefont{and}
  \bibinfo{author}{\bibfnamefont{E.}~\bibnamefont{Knill}},
  \bibinfo{journal}{Quantum Information and Computation}
  \textbf{\bibinfo{volume}{12}}, \bibinfo{pages}{553} (\bibinfo{year}{2012}).

\bibitem[{\citenamefont{Allan}(1966)}]{allan}
\bibinfo{author}{\bibfnamefont{D.}~\bibnamefont{Allan}},
  \bibinfo{journal}{Proceedings of the IEEE} \textbf{\bibinfo{volume}{54}},
  \bibinfo{pages}{221 } (\bibinfo{year}{1966}), ISSN \bibinfo{issn}{0018-9219}.

\bibitem[{\citenamefont{Riley}(2008)}]{riley:qc2008a}
\bibinfo{author}{\bibfnamefont{W.~J.} \bibnamefont{Riley}},
  \emph{\bibinfo{title}{Handbook of Frequency Stability Analysis}}, vol.
  \bibinfo{volume}{NIST Special Publication 1065} (\bibinfo{publisher}{NIST},
  \bibinfo{address}{Boulder, CO}, \bibinfo{year}{2008}).

\bibitem[{\citenamefont{Numata et~al.}(2004)\citenamefont{Numata, Kemery, and
  Camp}}]{optical1f}
\bibinfo{author}{\bibfnamefont{K.}~\bibnamefont{Numata}},
  \bibinfo{author}{\bibfnamefont{A.}~\bibnamefont{Kemery}}, \bibnamefont{and}
  \bibinfo{author}{\bibfnamefont{J.}~\bibnamefont{Camp}},
  \bibinfo{journal}{Phys. Rev. Lett.} \textbf{\bibinfo{volume}{93}},
  \bibinfo{pages}{250602} (\bibinfo{year}{2004}),
  \urlprefix\url{http://link.aps.org/doi/10.1103/PhysRevLett.93.250602}.

\bibitem[{\citenamefont{Fyodorov et~al.}(2009)\citenamefont{Fyodorov, Doussal,
  and Rosso}}]{1f}
\bibinfo{author}{\bibfnamefont{Y.~V.} \bibnamefont{Fyodorov}},
  \bibinfo{author}{\bibfnamefont{P.~L.} \bibnamefont{Doussal}},
  \bibnamefont{and} \bibinfo{author}{\bibfnamefont{A.}~\bibnamefont{Rosso}},
  \bibinfo{journal}{Journal of Statistical Mechanics: Theory and Experiment}
  \textbf{\bibinfo{volume}{2009}}, \bibinfo{pages}{P10005}
  (\bibinfo{year}{2009}),
  \urlprefix\url{http://stacks.iop.org/1742-5468/2009/i=10/a=P10005}.

\bibitem[{\citenamefont{Diddams et~al.}(2004)\citenamefont{Diddams, Bergquist,
  Jefferts, and Oates}}]{clockOverview}
\bibinfo{author}{\bibfnamefont{S.~A.} \bibnamefont{Diddams}},
  \bibinfo{author}{\bibfnamefont{J.~C.} \bibnamefont{Bergquist}},
  \bibinfo{author}{\bibfnamefont{S.~R.} \bibnamefont{Jefferts}},
  \bibnamefont{and} \bibinfo{author}{\bibfnamefont{C.~W.} \bibnamefont{Oates}},
  \bibinfo{journal}{Science} \textbf{\bibinfo{volume}{306}},
  \bibinfo{pages}{1318} (\bibinfo{year}{2004}),
  \eprint{http://www.sciencemag.org/content/306/5700/1318.full.pdf},
  \urlprefix\url{http://www.sciencemag.org/content/306/5700/1318.abstract}.

\bibitem[{\citenamefont{Chou et~al.}(2010)\citenamefont{Chou, Hume, Koelemeij,
  Wineland, and Rosenband}}]{PhysRevLett.104.070802}
\bibinfo{author}{\bibfnamefont{C.~W.} \bibnamefont{Chou}},
  \bibinfo{author}{\bibfnamefont{D.~B.} \bibnamefont{Hume}},
  \bibinfo{author}{\bibfnamefont{J.~C.~J.} \bibnamefont{Koelemeij}},
  \bibinfo{author}{\bibfnamefont{D.~J.} \bibnamefont{Wineland}},
  \bibnamefont{and}
  \bibinfo{author}{\bibfnamefont{T.}~\bibnamefont{Rosenband}},
  \bibinfo{journal}{Phys. Rev. Lett.} \textbf{\bibinfo{volume}{104}},
  \bibinfo{pages}{070802} (\bibinfo{year}{2010}),
  \urlprefix\url{http://link.aps.org/doi/10.1103/PhysRevLett.104.070802}.

\bibitem[{\citenamefont{Hanneke et~al.}(2009)\citenamefont{Hanneke, Home, Jost,
  Amini, Leibfried, and Wineland}}]{2QubitIon}
\bibinfo{author}{\bibfnamefont{D.}~\bibnamefont{Hanneke}},
  \bibinfo{author}{\bibfnamefont{J.~P.} \bibnamefont{Home}},
  \bibinfo{author}{\bibfnamefont{J.~D.} \bibnamefont{Jost}},
  \bibinfo{author}{\bibfnamefont{J.~M.} \bibnamefont{Amini}},
  \bibinfo{author}{\bibfnamefont{D.}~\bibnamefont{Leibfried}},
  \bibnamefont{and} \bibinfo{author}{\bibfnamefont{D.~J.}
  \bibnamefont{Wineland}}, \bibinfo{journal}{Nature Physics}
  \textbf{\bibinfo{volume}{6}}, \bibinfo{pages}{13} (\bibinfo{year}{2009}).

\bibitem[{\citenamefont{Eaton and Eaton}(1983)}]{eaton1983multivariate}
\bibinfo{author}{\bibfnamefont{M.~L.} \bibnamefont{Eaton}} \bibnamefont{and}
  \bibinfo{author}{\bibfnamefont{M.}~\bibnamefont{Eaton}},
  \emph{\bibinfo{title}{Multivariate statistics: a vector space approach}}
  (\bibinfo{publisher}{Wiley New York}, \bibinfo{year}{1983}).

\end{thebibliography}

\appendix

\section{Cost Function}
\label{app:A}

To simplify the notation, in the following theorem we write
$\theta=\theta_{n-1}$, $a=a_{n}$ and suppress the conditioning on
earlier measurement outcomes.

\begin{theorem}
\label{th:newCostFunction}
Consider a fixed interrogation duration $T$ and use measurement outcomes with
labels $a$ denoting arbitrary frequencies.  An ideal SDP with the cost
function $C = (\omega T - a)^2 + 2 ( \omega T - a )\mathbf{E}( \theta
- \mathbf{E}(\theta) | \omega )$ achieves the minimum expected
posterior variance increase of the cumulative phase $\Delta V=\sum_a
\mathbf{V}( \theta + \omega T | a ) p(a) - \mathbf{V}( \theta )$.
\end{theorem}

Ref.~\cite{firstClock} shows that it makes sense to talk about such an
ideal SDP, and that the objective values of its discretizations
converge to the ideal SDP's value. The discretization errors are well
behaved and can be effectively estimated.

\begin{proof}
  For now, we consider fixed algorithms $\mathcal{A}$ and do not
  identify measurement outcome labels with frequencies.  For clarity,
  we express integrals over measurement outcomes as discrete sums.
  Consider the expression for $\Delta V$ and expand it as follows:
  \begin{equation} 
    \begin{array}[b]{rl@{}l}
      \Delta V &=
      \rlap{$\displaystyle\sum_a \mathbf{V}( \theta + \omega T | a ) p(a) -  \mathbf{V}(\theta)$}&
      \\
      &=
      \displaystyle\sum_a \int \int &\left(\theta' + \omega' T - \mathbf{E}\left( \theta + \omega T | a \right) \right)^2 \\
      &&p( \theta', \omega' | a ) p( a ) d\omega' d\theta'  - \mathbf{V}(\theta).
    \end{array}
    \label{eq:posteriorVariance}
  \end{equation}
  Since, in general,
  \begin{equation} 
    \label{eq:minPosteriorVariance}
    \argmin_y \mathbf{E}\left(\left( X - y\right)^2 \right) =\mathbf{V}(X), \end{equation}
  we can rewrite Eq.~\eqref{eq:posteriorVariance} as
  \begin{equation} 
    \begin{array}[b]{rl@{}l}
      \Delta V &=
      \displaystyle\min_g \sum_a \int \int &\left( \theta' + \omega' T - g(a) \right)^2\\
      && p( \theta', \omega' | a ) p( a ) d\omega' d\theta' - \mathbf{V}(\theta), 
      \label{eq:firstTerm} 
    \end{array}
  \end{equation}
  where the minimum is over all functions $g$ of measurement outcomes.
  We can subtract any constant from inside the square of
  Eq.~\eqref{eq:firstTerm} without changing its value, as any constant
  shift will get absorbed in the minimum over $g$.  We choose to
  subtract the constant $\mathbf{E}(\theta)$, yielding
  \begin{equation} 
    \begin{array}[b]{rl@{}l}
      \Delta V &=\displaystyle
      \min_g \sum_a \int \int &\left( \theta' -\mathbf{E}\left(\theta\right) + \omega' T - g(a) \right)^2 \\
      &&p( \theta', \omega' | a ) p( a ) d\omega' d\theta' - \mathbf{V}(\theta).  
    \end{array}
    \label{eq:firstTermShifted1} 
  \end{equation}
  Expanding the square gives
  \begin{equation}
    \begin{array}[b]{rl@{}l}
      \Delta V &=\displaystyle
      \min_g \sum_a \int \int \Big(&\left(\theta' -\mathbf{E}\left(\theta\right)\right)^2\\
      &&{}+ \left(\omega' T - g(a) \right)^2\\
      &&{}+  2(\theta' - \mathbf{E}\left(\theta\right))\left( \omega' T - g(a) \right) \Big) \\
      &&p( \theta', \omega' | a ) p( a ) d\omega' d\theta' - \mathbf{V}(\theta).
    \end{array}
  \end{equation}
  Integrating out $\omega'$ in the first term gives a summand of
  $\mathbf{V}(\theta)$ that cancels the subtracted
  $\mathbf{V}(\theta)$.  We can then factor $p( \theta', \omega' | a )
  = p(\theta' | \omega', a )p(\omega' | a )$. We know that $\theta'$
  is conditionally independent of $a$ given $\omega'$, that is $p(
  \theta' | \omega', a ) = p( \theta' | \omega' )$, since $a$'s
  distribution is completely determined by $\omega'$ and the
  algorithm.  We can therefore rewrite
  Eq.~\eqref{eq:firstTermShifted1} as
  \begin{equation} \label{eq:difference2}
    \begin{array}[b]{rll@{}l}
      \Delta V &=&\displaystyle
      \min_g \sum_a \int& \bigg( \displaystyle\int\left(\omega' T - g(a) \right)^2 p(\theta' | \omega' ) d\theta'\\ 
      &\rlap{$\displaystyle{}+ 2\left(\omega' T - g(a) \right)\int\left( \theta' - \mathbf{E}\left(\theta\right)\bigg) p(\theta' | \omega' ) d\theta' \right)$}\\
      &&& p( \omega' | a ) p(a) d\omega'.
    \end{array}
  \end{equation}
  We carry out the integral over $\theta'$ and obtain
  \begin{equation}
    \begin{array}[b]{l}
      \Delta V =\displaystyle\min_g \sum_a \int \\
      \displaystyle \left( \left(\omega' T - g(a) \right)^2 
        + 2\left(\omega' T - g(a) \right)\mathbf{E}\left(\theta - \mathbf{E}\left(\theta\right) | \omega'\right)\right)\\
      \hphantom{\Delta V =\displaystyle\min_g \sum_a \int}
      p( \omega', a ) d\omega'.
    \end{array}
    \label{eq:th1finalintegral}
  \end{equation}
  Define $\bar C(g,\mathcal{A})$ to be the expression minimized over
  $g$ in this identity. It is of the form required by
  Eq.~\eqref{eq:clockCost} for the cost function $C$ of the theorem.
  Here, $\mathcal{A}$ denotes the previously implicit algorithm used
  for the interrogation.  The SDP for $C$ optimizes $\bar
  C(g,\mathcal{A})$ for a fixed $g$ over choices for
  $\mathcal{A}$. Its objective value is therefore an upper bound on
  $\Delta V$ for the algorithm found.
 
  Consider now the ideal SDP where the outcomes $a$ are arbitrary frequencies
  and $g(a)=a$. The optimization over $g$ is now redundant, because
  this SDP can realize any $\bar C(g,\mathcal{A})$ by relabeling the
  measurement outcomes. Thus, its objective value is the minimum
  variance increase.
\end{proof}

If we consider a discretized version of the SDP in the theorem with
fixed $g$, from
Eqs.~\eqref{eq:posteriorVariance},~\eqref{eq:minPosteriorVariance}
and~\eqref{eq:firstTerm} we deduce that $\Delta V$ for the SDP's
algorithm $\mathcal{A}$ can be computed by replacing $g$ with $g'$
defined by $g'(a) = \mathbf{E}(\theta + \omega T | a )$ in the
expression for $\bar C$. Since $\bar C(g',\mathcal{A})\leq \bar
C(g,\mathcal{A})$, one can re-evaluate the SDP with $g'$ in place of
$g$. Iterating this procedure in the limit yields an algorithm for
which $g=g'$. Whether the resulting algorithm achieves the optimal
$\Delta V$ may depend on the starting choices and the number of
measurement outcome labels. But the bounds on discretization error
from Ref.~\cite{firstClock} guarantee that the solution can be made
arbitrarily close to optimal.

Observe that the two terms of Eq.~\eqref{eq:th1finalintegral} resemble
$\mathbf{V}(\omega T)$ and $2\mathbf{Cov}( \omega T, \theta)$,
respectively, except that for the optimal choice of $g$, the offset
for $\omega T$ is not its mean. The dependence of the second term on
$\mathbf{E}(\theta|\omega)$ prevents the cost from being a
simple quadratic.

\section{Conditional Multivariate Gaussians}
\label{app:B}

For the noise models used here, the prior distribution $p(\omega_1,
\omega_2 \hdots \omega_N ) \equiv p( \boldsymbol{\omega_N})$ is a
multivariate Gaussian with means given by $\boldsymbol{ \mu } = (
\mathbf{E}(\omega_1 ), \mathbf{E}(\omega_2), \hdots
\mathbf{E}(\omega_N) )$ and covariance matrix $\boldsymbol{C}_{i,j} =
\mathbf{Cov}( \omega_i, \omega_j)$.  These means and covariances
completely characterize the distribution.  The clock updates require
computing $p( \omega_N | \boldsymbol{ \omega_{N-1} } )$. This
conditional probability distribution is also Gaussian and it suffices
to compute its mean and variance.  Denote the submatrix of
$\boldsymbol{C}$ containing rows $r$ through $s$ and columns $c$
through $d$ as $\boldsymbol{C}_{[r,s], [c, d]}$, and define subvectors
$\boldsymbol{\mu}_{[r,s]}$ of $\boldsymbol{\mu}$ in the same way.
The desired mean is given by \cite{eaton1983multivariate}
\begin{equation} 
\mu' = \mu_N + \boldsymbol{C}_{N,[1,N-1]} \boldsymbol{C}_{[1, N-1], [1, N-1]}^{-1}( \boldsymbol{f}  - \boldsymbol{\mu}_{[1, N-1]} ),
\end{equation}
and the variance by
\begin{equation}
C' = C_{N,N} - \boldsymbol{C}_{N,[1,N-1]} \boldsymbol{C}_{[1, N-1],[1, N-1]}^{-1} \boldsymbol{C}_{[1, N-1],N}.
\end{equation}

\section{Expectation Updates}
\label{app:C}

Before we can obtain the quantum algorithm for the next interrogation,
it is necessary to compute the parameters of the cost-function of
Eq.~\eqref{eq:costFunction}. These parameters depend on conditional
expectations of $\theta_{n}$. Because $\theta_{n}$ depends on all
frequencies since the clock was started, it is not clear how to
compute these expectations when the history is truncated to keep the
memory requirements manageable. Here we show that the relevant
expectations can be updated correctly with respect to the noise model
implied by the truncation strategy and without requiring additional
distributions to be maintained.

Truncation converts the ideal noise model into one with finite memory
as far as the frequencies $\omega_{n}$ are concerned. The prior
distribution for $\omega_{n+1}$ is computed taking into account only
its covariances with $\omega_{n},\ldots,\omega_{n-m+1}$, where the
history is truncated after $m$ interrogations. The truncated noise
model satisfies that $\omega_{n+1}$ is conditionally independent of
$\omega_{l}$ for $l<n-m+1$ given $\omega_{n},\ldots,\omega_{n-m+1}$.
Here we consider the more general situation, where the relevant state
of the oscillator after the $n$'th interrogation is parameterized by
$s_{n}$. For the truncated history and resulting noise models used
here, $s_{n}=(\omega_{n},\ldots,\omega_{n-m+1})$.  Given this setup
and the accordingly modified (though not ideal) noise model, we can
ensure that the distributions of $\omega_{n+k}$ ($k>0$) are
conditionally independent of $s_{n-l}$ ($l>0$) and $\theta_{n}$ given
$s_{n}$ and $\mathbf{a}_{n}$.  
We also assume that the values of all the relevant random variables
have been discretized, so that integrals are replaced by sums.

We now show how to keep track of the conditional moments $M_{k,n}=\mathbf{E}(
\theta_n^k | s_n, \boldsymbol{a_n} )$ for $k\leq K$ as we update the
various conditional distributions needed to compute priors and
posteriors. The cost function needed to optimize the $n{+}1$'th
interrogation requires the expectations $\mathbf{E}(\theta_{n}|
\omega_{n+1},\mathbf{a}_{n})$ and $\mathbf{E}(\theta_{n}|
\mathbf{a}_{n})$. The second can be obtained from the first by
integrating over $\omega_{n+1}$ with respect to the distribution
$p(\omega_{n+1}|\mathbf{a}_{n})=\sum_{s_{n}}p(\omega_{n+1}|s_{n},\mathbf{a}_{n})p(s_{n}|\mathbf{a}_{n})$. These
conditional distributions are available and updated by the protocol
after each interrogation. Given $M_{1,n}$, the first expectation can
be computed by setting $k=1$ in the following:
\begin{equation}
  \begin{array}[b]{l}
    \displaystyle\mathbf{E}(\theta_n^k|\omega_{n+1},\boldsymbol{a_n}) \\
    \displaystyle\quad{} = \sum_{s_n} \mathbf{E}(\theta_n^k|\omega_{n+1},s_n,\boldsymbol{a_n})p(s_n|\omega_{n+1},\boldsymbol{a_n}) \\
    \displaystyle\quad{}= \sum_{s_n,\theta_n} \theta_n^k
    p(\theta_n^k|\omega_{n+1},s_n,\boldsymbol{a_n})p(s_n|\omega_{n+1},\boldsymbol{a_n})  \\
    \displaystyle\quad{}= \sum_{s_n,\theta_n} \theta_n^k
    p(\theta_n^k|s_n,\boldsymbol{a_n})p(s_n|\omega_{n+1},\boldsymbol{a_n}) \notag \\
   \displaystyle\quad{} = \sum_{s_n} \mathbf{E}(\theta_n^k|s_n,\boldsymbol{a_n})p(s_n|\omega_{n+1},\boldsymbol{a_n}) \\
   \displaystyle\quad{} = \sum_{s_n} M_{k,n}p(s_n|\omega_{n+1},\boldsymbol{a_n}). 
  \end{array}
\end{equation}
In the third identity we applied the conditional independence of
$\omega_{n+1}$ and $\theta_n$ given $s_n$ and $\mathbf{a}_{n}$. The
factor in the last sum is determined by the noise model and is
available to the protocol.  We observe that the mean-square-errors
needed for evaluating protocol performance can be obtained from the
second moments ($k=2$) without the need for a full Monte Carlo
simulation.

For computing $M_{k,n+1}$ from the $M_{k',n}$, we are given $p( s_n |
\boldsymbol{a_n} )$ and can compute $p( \omega_{n+1}, s_n |
\boldsymbol{a_n} )$ and all derived conditionals and marginals.  At
this point we also know the outcome $a_{n+1}$.  Expand $M_{k,n+1}$ as
follows:
\begin{equation} \label{eq:E1}
  \begin{array}[b]{l}
    \displaystyle\mathbf{E}(\theta_{n+1}^k|s_{n+1},\boldsymbol{a_{n+1}}) \\
    \displaystyle\quad{} = \mathbf{E}\left(\left(\theta_n + \omega_{n+1}T_{n+1}\right)^k|s_{n+1},\boldsymbol{a_{n+1}}\right) \\
    \displaystyle\quad{} = \sum_{j=0}^k \mathbf{E}\left( {k\choose j}
      T_{n+1}^j \omega_{n+1}^j\theta_n^{k-j} | s_{n+1},\boldsymbol{a_{n+1}}\right).
  \end{array}
\end{equation}
To evaluate the $j$'th term of this sum we can compute
\begin{equation} \label{eq:E2}
  \begin{array}[b]{l}
    \displaystyle\mathbf{E}(\omega_{n+1}^j T^j_{n+1}\theta_n^{k-j}|s_{n+1},\boldsymbol{a_{n+1}}) \\
    \displaystyle\quad{} = \sum_{\omega_{n+1}}\omega_{n+1}^j  T^j_{n+1} \mathbf{E}(\theta_n^{k-j}|\omega_{n+1},s_{n+1},\boldsymbol{a_{n+1}})\\
    \displaystyle\hphantom{\quad{}={}\sum_{\omega_{n+1}}\omega_{n+1}^j  T^j_{n+1} }
     \times p(\omega_{n+1}|s_{n+1},\boldsymbol{a_{n+1}}).
  \end{array}
\end{equation}
We have that $\theta_n$ is conditionally independent of $\omega_{n+1}$, $s_{n+1}$,
and $a_{n+1}$ given $s_n$ and $\boldsymbol{a}_n$.  Therefore,
\begin{equation} \label{eq:E3}
  \begin{array}[b]{l}
    \displaystyle\mathbf{E}(\theta_n^{k-j}|\omega_{n+1},s_{n+1},\boldsymbol{a_{n+1}})  \\
    \displaystyle\quad{} = \sum_{s_n}\mathbf{E}(\theta_n^{k-j}|\omega_{n+1},s_{n+1},s_n,\boldsymbol{a_{n+1}})\\ 
    \displaystyle\phantom{\quad{} = \sum_{s_n}\mathbf{E}(\theta_n^{k-j}|}
    \times p(s_n|\omega_{n+1},s_{n+1},\boldsymbol{a_{n+1}}) \\

    \displaystyle\quad{}= \sum_{s_n}\mathbf{E}(\theta_n^{k-j}|s_n, \boldsymbol{a_n}) p(s_n|
    \omega_{n+1},s_{n+1},\boldsymbol{a_{n+1}})\\
    \displaystyle\quad{}= \sum_{s_n}M_{k-j,n}p(s_n|
    \omega_{n+1},s_{n+1},\boldsymbol{a_{n+1}}).
  \end{array}
\end{equation}
The last factor in the summand is determined by the noise model and
the algorithm used for the $n+1$'th interrogation. It is can therefore
be computed from the posteriors maintained by the protocol.

\section{Timing Error Suppression}
\label{app:D}

We describe methods for suppressing the errors due to differences
between $T_{n}$ and the true interrogation duration determined from the
external oscillator, and the errors from non-instantaneous state
preparation and measurement.  We argue that with proper implementation
design, uncertainties in these durations are a small fraction of the
intended interrogation duration $T$, which results in relatively small
biases when inferring external oscillator frequencies.  This requires
that the change in actual interrogation duration does not significantly
affect the noise accumulated according to the noise model and that
there is little change in the conditional probability distributions of
the measurement outcomes given the true frequency $\omega$ of the
oscillator for the duration of the interrogation. Relative to the
accumulated noise for the total interrogation duration $T$, the
contribution associated with differences between $T$ and the effective
interrogation duration $T'$ relates to $(T-T')/T$ with a corresponding
small effect on the clock.  The conditional probability distributions
are determined by the measurement procedure, which our protocol
specifies in the frame of the atomic standard at the end of the
interrogation period.  The implementation must use the frame of the
external oscillator instead, so the measurement is adjusted for the
experimenter's best estimate of the relative phases.  Because the
oscillator is classical, the experimenter has access to the absolute
oscillator phase $\phi$ relative to the beginning of the
interrogation. This phase relates to the true time difference $s$
according to $\phi=(\omega+\Omega) s$, where, for current purposes,
$\omega$ is the true average frequency deviation from $0$ to $s$, and
$\Omega$ is the (unchanging) frequency of the atomic standard.  The
quantum algorithm obtained in the procedure expects that the
measurement is at time $s=T$ and the relative phase of the atomic
standard compared to the oscillator at this time is $-\omega T$. For
$s\not=T$ the actual relative phase is $-\omega s$, which can be
substantially different if the oscillator has drifted and the
measurement duration is non-negligible.  The experimenter can compensate
for this issue by modifying the measurement phase in time according to
the best estimate $\omega^{*}$ of $\omega$. If $s$ is known, a good
compensating phase is $\omega^{*}(s-T)$, and with this compensation,
the phase error is reduced to $\epsilon=(\omega^{*}-\omega)(s-T)$, so
the measurement is not sensitive to long term drift of the oscillator.
Note that this procedure is equivalent to offsetting the oscillator
frequency by $-\omega^{*}$, which corresponds to the standard practice
of controlling the oscillator to stay close to the atomic standard.
To avoid having to modify our noise model and representations of
probability distributions, we find it convenient to perform this
control in software instead.

The above compensating phase $\omega^{*}(s-T)$ cannot be used directly
since $s$ is not known. The experimenter's best estimate for $s$ is
$s^{*}=\phi/(\omega^{*}+\Omega)$.  If this is used, the compensating
phase at the time of measurement is $\omega^{*}(s^{*}-T)$, or $\phi\,
\omega^{*}/(\omega^{*}+\Omega) - \omega^{*}T$ in terms of the external
oscillator phase $\phi$.  The phase error is
\begin{align}
\epsilon  &= -\omega T -(-\omega s + \omega^{*}(s^{*}-T)) \nonumber\\
  &= (\omega^{*}-\omega) (T-s)
       +\omega^{*}(s-s^{*}) \nonumber\\
  &= (\omega^{*}-\omega) (T-s)
       +\omega^{*}s\left(1-\frac{\omega+\Omega}{\omega^{*}+\Omega}\right) \nonumber\\
  &= (\omega^{*}-\omega) \left((T-s) + \frac{\omega^{*}}{\omega^{*}+\Omega}s\right).
\end{align}
To avoid phase slip, it is necessary to choose $T$ such that
$(\omega^{*}-\omega)T <\pi$ with high probability.  If this inequality
holds, then the phase error is bounded by 
\begin{equation}
E_{1}=\pi|T-s|/T + \pi
(s/T)(\omega^{*}/(\omega^{*}+\Omega)), 
\label{eq:errorbnd1}
\end{equation}
both of which are expected to
be small. Furthermore, even without controlling the oscillator to
avoid large excursions, we expect the first term to dominate.  

To avoid problems from finite preparation and measurement durations, the
compensating phase must be applied continuously in time.  A direct way
to do this is by providing an auxilliary oscillator locked to the
external oscillator and offset by $-\omega^{*}$. Specifically, the
phase of the auxilliary oscillator is given by
$-\phi\omega^{*}/(\omega^{*}+\Omega)$ with respect to the phase $\phi$
of the external oscillator. For state preparation, operations applied
to the atom have phase $0$ with respect to the auxilliary oscillator.
For measurement, the phase with respect to the auxilliary oscillator
and added to the phases of the measurement computed by the SDP is
given by $T\omega^{*}$. The measurement period is centered around the
time when the phase of the external oscillator is
$\phi=T(\omega^{*}+\Omega)$.  With this procedure, the error due to
preparation and measurement durations of order $\Delta T$ is directly
related to the phase error due to non-ideal true measurement durations of
the same order.

Large excursions of $\omega$ compared to $\Omega$ are not normally
expected. Nevertheless, it may be desirable to eliminate the second
term contributing to the phase error in Eq.~\eqref{eq:errorbnd1}.  For
this purpose, one can modify the SDP used to compute the optimal
protocol. If the experimenter determines the end of the interrogation
according to $\phi_{T}=T(\omega^{*}+\Omega)$, the true time at the end
is
$\phi_{T}/(\omega+\Omega)=T(\omega^{*}+\Omega)/(\omega+\Omega)$. The
relative phase is $\rho(\omega)T=\omega T
(\omega^{*}+\Omega)/(\omega+\Omega)$ instead of $\omega T$.  This
changes the relationship between $\omega$ and the phase of the
interrogation unitary, since the construction of the SDP as given in
the text assumes that the accumulated phase difference is $-\omega
T$. The modified phase difference can be accommodated by a
re-parameterization of the frequencies in the SDP.  This is
accomplished by defining $\omega'=\rho(\omega)$, computing the prior
needed by the SDP for $\omega'$ from that for $\omega$ accordingly and
using the cost function $C'$ defined by $C'(\omega',a) =
C(\rho^{-1}(\omega'),a)$ in Eq.~\eqref{eq:clockCost}.  The
continuously applied measurement phase compensation to account for the
non-instantaneous and inexact measurement is then given by
$(\phi-\phi_{T})\omega^{*}/(\omega^{*}+\Omega)$ as a function of the
oscillator phase $\phi$, which is identical to that given by the
earlier method that simulates a locked oscillator, previously
expressed as $\phi\omega^{*}/(\omega^{*}+\Omega)-\omega^{*}T$.  With
the reparametrized cost function and this phase compensation, the
remaining phase error compared to protocol expectation is given by
\begin{align}
  \epsilon &=
  (\phi-\phi_{T})\left(\frac{\omega}{\omega+\Omega}-\frac{\omega^{*}}{\omega^{*}+\Omega}\right) \nonumber\\
  &= (\omega-\omega^{*})\frac{\phi-\phi_{T}}{\omega+\Omega}\;\frac{\Omega}{\omega^{*}+\Omega}.
\end{align}
To explicitly compare this to the earlier error, note that $s-T=
(\phi-\phi_{T})/(\omega+\Omega)$ corresponds to the difference between
the ideal interrogation duration and the implemented one, and
$(\omega-\omega^{*})T < \pi$, so the error is bounded by $\pi (|T-s
|/T)\Omega/(\omega^{*}+\Omega)$. This is expected to be small but
shows that low absolute oscillator frequencies require correspondingly
more precise interrogation durations.  Note that in general, low-frequency
oscillators do not make good clocks and real noise models are not
frequency independent.

\end{document}